\documentclass[twocolumn,showpacs,preprintnumbers,amsmath,amssymb]{revtex4}

\usepackage{amsmath}
\usepackage{amsfonts}
\usepackage{amssymb}
\usepackage{graphicx}
\usepackage{subfigure}
\usepackage{graphicx}
\usepackage{dcolumn}
\usepackage{bm}

\begin{document}


\title{Tunneling of conduction band electrons driven by a laser field in a double quantum dot: An open systems approach}

\author{B. Ahmadi}
\author{S. Salimi}
\email{shsalimi@uok.ac.ir}
\author{A. S. Khorashad}
\email{a.sorouri@uok.ac.ir}
\affiliation{ %
Department of Physics, Faculty of Science, University of Kurdistan,  Sanandaj, Iran\\
}%


\date{\today}

\def\br{\biggr}
\def\bl{\biggl}
\def\Br{\Biggr}
\def\Bl{\Biggl}
\def\be\begin{equation}
 \def\ee{\end{equation}}
\def\bea{\begin{eqnarray}}
\def\eea{\end{eqnarray}}
\def\f{\frac}
\def\n{\nonumber}
\def\l{\label}
\begin{abstract}
In this paper, we investigate tunneling of conduction band electrons
in a system of an asymmetric double quantum dot which interacts with
an environment. First, we consider the case in which the system only
interacts with the environment and demonstrate that as time goes to
infinity they both reach an equilibrium, which is expected, and
there is always a maximum and minimum for the populations of the
states of the system. Then we investigate the case in which an
external resonant optical pulse (a laser) is applied to the system
interacting with the environment. However, in this case for
different intensities we have different populations of the states in
equilibrium and as the intensity of the laser gets stronger, the
populations of the states in equilibrium approach the same constant.
\end{abstract}

\keywords{Suggested keywords}
\maketitle
\newpage
\section{Introduction}
Developments in semiconductors have let us design quantum dots
(QDs) \cite{1, 2, 3}.
 In these systems, few electrons may be confined in a nanometric space, which would be like atoms where few electrons are confined.
A usual way to construct QDs is to trap a two-dimensional electronic
gas in a semiconductor. This process of confinement creates a
bowl-like potential in which conduction band electrons are trapped \cite{4}.
This is called a quantum dot. Quantum dots have let us make
nanostructure devices which are governed by quantum mechanics rules
\cite{5, 6, 7, 8, 9}.
 It is predicted that in the future quantum dots will have numerous
applications in industry like in memory chips \cite{10}, quantum
computation \cite{11, 12}, and quantum cryptography \cite{13}. In
this work, we investigate a three-level system made up of a double
quantum dot. J. M. Villas-Boas et. al \cite{14} have investigated
such a system, but, what we do is totally different. In fact, here,
a much more real system is considered, which interacts with the
environment. The system is assumed to be a {Markovian} one
\cite{15}. The system and the bath can exchange energy, and this
leads to dissipation, and fluctuations. Since the system is open and
Markovian, the quantum optical master equation is used to
investigate its evolution \cite{15}. In the next section we let the
system interact with the bath and write its equation of motion, then
obtain and plot the results. In the third section we apply a laser
to the system interacting with bath and then again write the
appropriate equation of motion and plot the results.
\section{System interacting with the environment}
The mutual influence of a system, with a few degrees of freedom, and
a heat bath, with many degrees of freedom, on each other is the
central concept in the physics of noise from both quantum and
classical point of view. Here, the system under consideration is a
QD-molecule formed from an asymmetric double QD system whose dots
are coupled through tunneling and interacts with the bath. Such a
system can be made by self-assembled dot growth technology
\cite{16}. The gap between energy levels of the right dot is much
bigger than that of the left dot. Thus the energy levels are out of
resonance, which means electron tunneling is weak (see Fig. 1(a)).
But if we apply a voltage gate, the conduction band levels of the
two dots get closer to resonance, which means their coupling
increases, while the valence band levels become more off-resonant,
meaning that their coupling becomes weak in a way that the
probability of tunneling of the hole (left from the transition of
the electron to the conduction band in the right dot) to the valence
band of the right dot becomes very small (see Fig. 1(b)). Since the
gap between the conduction band and the valence band is much bigger
than that of the left dot we can assume that the bath is unable to
excite the electrons of the valence band of the right dot (this is
like a two-level atom interacting with the bath; although the system
is a three-level system, the two of them are degenerate). It is also
assumed that the system is initially in the state that there is no
electron in the conduction band of the left dot (the state
$|0\rangle$). Therefore when the system interacts with the bath,
this interaction causes an electron to leave the valence band and
jump into the conduction band and it leaves a hole in the valence
band (the state $|1\rangle$) and then since the electron tunneling
couples the conduction bands of the two dots, the electron
\begin{figure}[h]
\centering
\includegraphics[width=6.4cm]{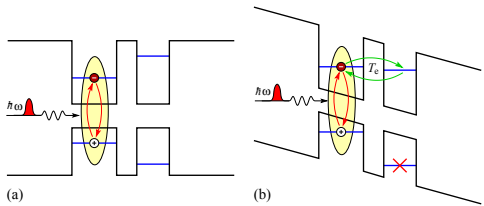}
\caption{(Color online) Schematic band structure of the system. (a)
without a gate voltage. (b) with applied gate voltage.}
\label{fig:Fig1.eps}
\end{figure}
can tunnel to the conduction band of the right dot, (the state
$|2\rangle$).In this section we address the case in which the system
only interacts with the bath and we see that the system and the bath
reach an equilibrium as it is expected, but there would be some
noticeable results about this equilibrium. The Hamiltonian of the
total system can be written as
\begin{equation}\label{1}
H_{tot}=H_S+H_B+H_I,
\end{equation}
where $H_S$ is the Hamiltonian of the system and $H_B$ is the
Hamiltonian of the bath and $H_I$ is the Hamiltonian describing the
interaction between the system and the bath. The Hamiltonian of the
uncoupled QD-molecule may be written in the form
\begin{equation}\label{2}
H_S=\sum_i\varepsilon_i|i\rangle\langle i|+T_e(|1\rangle\langle2|+|2\rangle\langle1|),
\end{equation}
where $T_e$ is the electron tunneling matrix
element which couples the conduction bands of the two dots.
The heat bath which is taken to be a free quantized radiation field (with an infinite number of degrees of freedom)
 will be represented by the Hamiltonian (subtracting the zero point energy)\cite{17},
\begin{equation}\label{3}
H_B=\sum_j\omega_jb_j^\dagger b_j,
\end{equation}
where $b_j^\dagger$ and $b_j$ are the creation and annihilation
operators of the field modes labeled by j, respectively. In fact a
field interpretation of the heat bath can almost always be possible.
When viewed from the position of the system (like an atom or our
QD-molecule here) there are two kinds of field modes, incoming and
outgoing. The incoming modes affect the system, whereas the outgoing
modes are produced by the system, they affect the system because
they carry away energy, and thus give rise to damping. Incoming
modes give the energy to the system, which is carried away by the
outgoing modes. Thus the sort of damping produced is radiation
damping. This view is true whenever a field interpretation of the
bath is possible, and this is almost always possible \cite{18}.
Finally, we assume the interaction Hamiltonian to be given in the
dipole approximation by
\begin{equation}\label{4}
H_I=-\textbf{D}.\textbf{E},
\end{equation}
where \textbf{D} is the dipole operator of the system under
consideration and \textbf{E} is the electric field operator in the
Schr\"{o}dinger picture \cite{15} which couples the states
$|0\rangle$ and $|1\rangle$,
\begin{equation}\label{5}
\textbf{E}=i\sum_k\sum_{\lambda=1,2}\sqrt{\dfrac{2\pi \hslash\omega_k}{V}}\textbf{e}_\lambda(k)\big(b_\lambda(k)-b^\dagger_\lambda(k)\big).
\end{equation}
The system is completely similar to a three-level system, therefore we expect to
have 6 Lindbland operators, two operators for each two levels
\cite{15}. But in our system the direct transitions $|0\rangle$
$\rightleftarrows$ $|2\rangle$ are neglected, as the dipole moment
for that spatially indirect exciton will be vanishingly small. We
thus have four Lindbland operators. For an arbitrary system
interacting with the heat bath the quantum optical master equation in the Lindbland form would be
\cite{15}
\begin{eqnarray}\label{6}\nonumber
\dot{\rho}(t)&=&\sum_{\omega>0}\frac{4\omega^3}{3\hslash c^3}\big(N(\omega)+1\big)\big(\textbf{A}(\omega)\rho(t)\textbf{A}^\dag(\omega)\\ \nonumber
&-&\frac{1}{2}[\textbf{A}^\dag(\omega)\textbf{A}(\omega),\rho(t)]_{+}\big)\\ \nonumber
&+&\sum_{\omega>0}\frac{4\omega^3}{3\hslash c^3} N(\omega)\big(\textbf{A}^\dag(\omega) \rho(t)\textbf{A}(\omega)\\
&-&\frac{1}{2} [\textbf{A}(\omega)\textbf{A}^\dag(\omega),\rho(t)]_{+}\big)
-\frac{i}{\hslash}[H_{LS},\rho_S(t)],
\end{eqnarray}
where
\begin{equation}\label{7}
N(\omega)=\frac{1}{\exp(\beta\hbar\omega)-1}
\end{equation}
is the Planck distribution, which is the average number of photons in a mode with frequency $\omega$ and $\textbf{A}(\omega)$ is called Lindbland operator and the last term is called the Lamb and Stark shift contribution. The terms $4\omega^3/3\hslash c^3(N(\omega)+1)$ and $4\omega^3/3\hslash c^3N(\omega)$ are the transition rates at which spontaneous and thermally induced emission and thermally induced absorption processes occur, respectively \cite{15}. It should be mentioned that the quantum optical master equation is a Markovian master equation. Markovian behavior seems reasonable on physical grounds. Potentially, the system may depend on its past history. If, however, the bath is a large system maintained in thermal equilibrium, we do not expect it to preserve the minor changes brought by its interaction with the system for a very long time; not for long enough to significantly affect the future evolution of the system \cite{19}. The interaction between matter and electromagnetic radiation is a typical and appropriate candidate for the application of quantum Markovian master equations \cite{18, 20}.
Lindbland operators are obtained by decomposing the dipole operator \textbf{D} into eigenvectors of $H_S$,

\begin{equation}\label{8}
\textbf{A}(\omega_{01})=\langle0|\textbf{D}|1\rangle|0\rangle\langle1|\nonumber,
\end{equation}
\begin{equation}\label{9}
\textbf{A}^\dag(\omega_{01})=\langle1|\textbf{D}|0\rangle|1\rangle\langle0|\nonumber,
\end{equation}
 \begin{equation}\label{10}
\textbf{A}(\omega_{12})=\langle1|\textbf{D}|2\rangle|1\rangle\langle2|\nonumber,
\end{equation}
 \begin{equation}\label{11}
\textbf{A}^\dag(\omega_{12})=\langle2|\textbf{D}|1\rangle|1\rangle\langle2|\nonumber.
\end{equation}
Now we define
\begin{equation}\label{12}
\sigma_{-01}=|0\rangle\langle1|\nonumber,
\end{equation}

\begin{equation}\label{13}
\sigma_{+01}=|1\rangle\langle0|\nonumber,
\end{equation}
\begin{equation}\label{14}
\sigma_{-12}=|1\rangle\langle2|\nonumber,
\end{equation}

\begin{equation}\label{15}
\sigma_{+12}=|2\rangle\langle1|\nonumber,
\end{equation}
and
\begin{equation}\label{16}
\textbf{d}_{01}=\langle0|\textbf{D}|1\rangle\nonumber,
\end{equation}

\begin{equation}\label{17}
\textbf{d}^*_{01}=\langle1|\textbf{D}|0\rangle\nonumber,
\end{equation}
\begin{equation}\label{18}
\textbf{d}_{12}=\langle 1|\textbf{D}|2\rangle\nonumber,
\end{equation}
\begin{equation}\label{19}
\textbf{d}^*_{12}=\langle2|\textbf{D}|1\rangle\nonumber.
\end{equation}
Therefore according to these definitions we have
\begin{equation}\label{20}
\textbf{A}(\omega_{01})=\textbf{d}_{01}\sigma_{-01}\nonumber,
\end{equation}
 \begin{equation}\label{21}
\textbf{A}^\dag(\omega_{01})=\textbf{d}^*_{01}\sigma_{+01}\nonumber,
\end{equation}
\begin{equation}\label{22}
\textbf{A}(\omega_{12})=\textbf{d}_{12}\sigma_{-12}\nonumber,
\end{equation}
 \begin{equation}\label{23}
\textbf{A}^\dag(\omega_{12})=\textbf{d}^*_{12}\sigma_{+12}\nonumber.
\end{equation}
Substituting these Lindbland operators into Eq.(\ref{6}) and neglecting the Lamb and Stark shift contribution \cite{15}, the quantum optical master equation can be written in the form
\begin{eqnarray}\label{24}\nonumber
\dot{\rho}(t)&=&\gamma_{01}\big(N(\omega_{01})+1\big)\big(\sigma_{-01}\rho(t)\sigma_{+01}\\
\nonumber
&-&\frac{1}{2}[\sigma_{+01}\sigma_{-01},\rho(t)]_{+}\big)\\
\nonumber &+&\gamma_{01} N(\omega_{01})\big(\sigma_{+01}
\rho(t)\sigma_{-01}-\frac{1}{2}
[\sigma_{-01}\sigma_{+01},\rho(t)\big]_{+}) \\ \nonumber
&+&\gamma_{12}\big(N(\omega_{12})+1\big)\big(\sigma_{-12}\rho(t)\sigma_{+12}\\
\nonumber
&-&\frac{1}{2} [\sigma_{+12}\sigma_{-12},\rho(t)]_{+}\big)+\gamma_{12} N(\omega_{12})\big(\sigma_{+12} \rho(t)\sigma_{-12}\\
&-&\frac{1}{2} [\sigma_{-12}\sigma_{+12},\rho(t)]_{+}\big),
\end{eqnarray}
where
\begin{equation}\label{25}
\gamma_{ij}=\frac{4\omega^3 |\textbf{d}_{ij}|^2}{3\hslash c^3}.
\end{equation}
We want to investigate the case in which the levels $|1\rangle$ and $|2\rangle$ are in resonance. But the problem is that if we let $\omega_{12}$ be zero, the terms containing $\omega_{12}$ vanish, and this seems like we have a two-level system which is obviously not the case here. Thus to deal with this problem we do the following. Since $T_e$ is nonzero, therefore the transition rate from $|1\rangle$ to $|2\rangle$ must not vanish. Hence it stands to reason if we add a constant n, proportional to $T_e$, to these transition rates.
Therefore for the resonance case ($\omega_{12}=0$), Eq.(\ref{24}) reads
\begin{eqnarray}\label{26}\nonumber
\dot{\rho}(t)&=&\gamma_{01}\big(N(\omega_{01})+1\big)\big(\sigma_{-01}\rho(t)\sigma_{+01}\\ \nonumber
&-&\frac{1}{2}[\sigma_{+01}\sigma_{-01},\rho(t)]_{+}\big)\\ \nonumber
&+&\gamma_{01} N(\omega_{01})\big(\sigma_{+01} \rho(t)\sigma_{-01}-\frac{1}{2} [\sigma_{-01}\sigma_{+01},\rho(t)]_{+}\big) \\ \nonumber
&+&n\big(\sigma_{-12}\rho(t)\sigma_{+12}-\frac{1}{2} [\sigma_{+12}\sigma_{-12},\rho(t)]_{+}\big)\\
&+&n\big(\sigma_{+12} \rho(t)\sigma_{-12}-\frac{1}{2} [\sigma_{-12}\sigma_{+12},\rho(t)]_{+}\big).
\end{eqnarray}
The matrix elements $\rho_{00}$, $\rho_{11}$ and $\rho_{22}$ are the populations of the states $|0\rangle$, $|1\rangle$ and $|2\rangle$, respectively. The off-diagonal elements $\rho_{ij}$ ($i\neq j$) are the coherences given by the expectation values of the raising and lowering operators $\sigma_{\pm}$.
The differential equations satisfied by $\rho_{00}$ and $\rho_{11}$ and $\rho_{22}$ are
\begin{equation}\label{27}
\dot{\rho}_{00}(t)=l\rho_{11}(t)-m\rho_{00}(t),
\end{equation}
\begin{equation}\label{28}
\dot{\rho}_{11}(t)=-(l+n)\rho_{00}(t)+m\rho_{11}(t)+n\rho_{22}(t),
\end{equation}
\begin{equation}\label{29}
\dot{\rho}_{22}(t)=n\rho_{11}(t)-n\rho_{22}(t),
\end{equation}
where $l$ equals $\gamma_{01}(N(\omega_{01})+1)$ and $m$ equals $\gamma_{01}N(\omega_{01})$.
Assuming that the system is initially in the state $|0\rangle$, one can solve these differential equations to obtain
\begin{equation}\label{30}
\rho_{00}(t)=\frac{l}{2m+l}+B\exp(\lambda_{0}t)+C\exp(\lambda_{1}t),
\end{equation}
\begin{eqnarray}\label{31}\nonumber
\rho_{11}(t)&=&\frac{m}{2m+l}+\frac{(\lambda_{0}+m)B}{l}\exp(\lambda_{0}t)\\
&+&\frac{(\lambda_{1}+m)C}{l}\exp(\lambda_{1}t),
\end{eqnarray}
\begin{eqnarray}\label{32}\nonumber
\rho_{22}(t)&=&\frac{m}{2m+l}-\big(B+\frac{(\lambda_{0}+m)B}{l}\big)\exp(\lambda_{0}t)\\
&-&\big(C+\frac{(\lambda_{1}+m)C}{l}\big)\exp(\lambda_{1}t),
\end{eqnarray}
where B and C are constants and
\begin{equation}\label{33}
\lambda_{0}=\frac{-(l+m+2n)-\sqrt{(l+m+2n)^2-4(ln+2mn)}}{2},
\end{equation}
\begin{equation}\label{34}
\lambda_{1}=\frac{-(l+m+2n)+\sqrt{(l+m+2n)^2-4(ln+2mn)}}{2}.
\end{equation}
B and C can be determined from the initial condition. The terms
including these constants vanish as time goes to infinity (See
appendix (A)). It is clear that as time goes to infinity $\rho_{00}$
decreases exponentially to a constant which is $l/(2m+l)$ and both
$\rho_{11}$ and $\rho_{22}$ increase from zero to the same constant
$m/(2m+l)$. In other words, in the limit of $t\rightarrow\infty$ the
solutions become stationary and this means that the system and the
environment are in equilibrium. What is interesting is that the
stationary values depend on $l$ and $m$ but not on $n$. $n$ just
determines how fast $\rho_{22}(t)$ reaches its stationary value. The
other interesting point is that the stationary value of $\rho_{11}$
and $\rho_{22}$ is always less than $0.33$ in the absence of an
external driving resonant optical pulse. In the next section, we
investigate the above-mentioned total system in the presence of a
laser and determine the condition under which $\rho_{11}$ and
$\rho_{22}$ can reach this maximum. Since the transitions
$|0\rangle$ $\rightleftarrows$ $|2\rangle$ are neglected, the system
takes on the state $|1\rangle$ before the $|2\rangle$, thus
$\rho_{11}$ reaches this constant faster than $\rho_{22}$ (see Figs.
$2-4$).
\section{System driven by an external resonant optical pulse (a laser)}
In this section we address the case in which the transition
from $|0\rangle$ to $|1\rangle$ is driven by an external laser on resonance.
 The Hamiltonian in this case can be written as
\begin{equation}\label{35}
H_{tot}=H_S+H_B+H_I+H_L,
\end{equation}
where $H_L$ is the Hamiltonian of the external resonant optical
pulse. Invoking the dipole approximation we obtain the following
interaction picture Hamiltonian describing the interaction of the
system with the driving mode:
\begin{equation}\label{36}
H_{L}(t)=-\textbf{E}_L(t).\textbf{D}(t),
\end{equation}
where
\begin{equation}\label{37}
\textbf{E}_L(t)=\textbf{e}\exp(-i\omega_{0}t)+\textbf{e}^*\exp(i\omega_{0}t),
\end{equation}
is the electric field strength of the driving mode. Since the laser
is on resonance with the transition from $|0\rangle$ to $|1\rangle$
therefore $\omega_{0}=\omega_{01}$. Now we can write $H_L(t)$ in the
rotating wave approximation as follows,
\begin{equation}\label{38}
H_{L}=-\frac{\Omega}{2}(\sigma_{+01}+\sigma_{-01}),
\end{equation}
where the product
\begin{equation}\label{39}
\Omega=2\textbf{e}.\textbf{d}^*_{01}
\end{equation}
is refereed to as the Rabi frequency.
Now neglecting the Lamb and Stark shift contribution again, the quantum optical master equation reads \cite{15}
\begin{eqnarray}\label{40}\nonumber
\dot{\rho}(t)&=&-\frac{\Omega}{2}[\sigma_{+01}+\sigma_{-01},\rho(t)]_{+}\\
\nonumber
&+&\gamma_{01}\big(N(\omega_{01})+1\big)\big(\sigma_{-01}\rho(t)\sigma_{+01}\\
\nonumber
&-&\frac{1}{2}[\sigma_{+01}\sigma_{-01},\rho(t)]_{+}\big)\\
\nonumber &+&\gamma_{01} N(\omega_{01})\big(\sigma_{+01}
\rho(t)\sigma_{-01}-\frac{1}{2}
[\sigma_{-01}\sigma_{+01},\rho(t)\big]_{+}) \\ \nonumber
&+&n\big(\sigma_{-12}\rho(t)\sigma_{+12}-\frac{1}{2} [\sigma_{+12}\sigma_{-12},\rho(t)]_{+}\big)\\
&+&n\big(\sigma_{+12} \rho(t)\sigma_{-12}-\frac{1}{2} [\sigma_{-12}\sigma_{+12},\rho(t)]_{+}\big).
\end{eqnarray}
Since for this case the elements of the density matrix are complex,
we have $18$ coupled differential equations. Diagonal elements are
coupled with the off-diagonal ones. Numerically solving these
equations we find that when time goes to infinity the results become
stationary (see appendix A). But for different amounts of
$2l/\Omega$, relaxation times and the number of oscillations are
different. The smaller $2l/\Omega$ is, the more the number of
oscillations and the longer relaxation time will be. This is clearly
due to the fact that when the intensity of the laser increases, the
laser drives more strongly the system to oscillate with the
frequency $\omega$ which is the frequency of the laser. From now on
we set $\Omega/2$ equal to $p$ for simplicity. The results are
plotted in figures $(5-13)$. We have not plotted $\rho_{22}(t)$ for
$l/p=1$, $l/p=1/2$ and $l/p=1/10$ because its evolution is very
similar to $\rho_{22}(t)$ when $l/p=2$ (i.e it exponentially
increases from zero to a constant and the only difference is that
these constants are different from each other for different values
of $l/p$ and it approaches $1/3$ as the intensity of the laser
increases.).
\section{conclusion}
We investigated a system of a quantum dot molecule and saw that
regardless of applying an external resonant optical pulse, as
$t\rightarrow\infty$ the system and the environment always reaches
an equilibrium, which is obvious and expected for a system
interacting with an environment. In equilibrium the populations
become stationary but as we see from Fig. 2 to Fig. 13 as $l/p$
decreases, a decrease in the stationary value of $\rho_{00}(t)$
begins and conversely an increase in those of $\rho_{11}(t)$ and
$\rho_{22}(t)$. But the most important point is that as $l/p$
decreases the stationary values of $\rho_{00}(t)$, $\rho_{11}(t)$
and $\rho_{22}(t)$ approach the same constant which is $1/3$. This
means that the three populations become equal in equilibrium when
$p\rightarrow\infty$. The reason is that the external laser causes
the system to leave the state $|0\rangle$ and jump to the state
$|1\rangle$, therefore the probability of finding the system in the
state $|0\rangle$ decreases and accordingly the probability of
finding the system in the state $|1\rangle$ and $|2\rangle$
increases. This in turn leads to a decrease in the population of the
state $|0\rangle$ and consequently a decrease in the stationary
value of $\rho_{00}(t)$ in equilibrium. This decrease continues as
the intensity of the laser increases until the stationary value of
$\rho_{00}(t)$ reaches $0.33$ and then it stops decreasing, i.e the
external laser causes the stationary value of $\rho_{00}(t)$ to
encounter a decrease to the final value $1/3$. But the stationary
values of $\rho_{11}(t)$ and $\rho_{22}(t)$ encounter an increase to
the final stationary value $1/3$ as the laser is applied. In other
words, as we increase the intensity of the laser, as a tunneling
controlling parameter, from zero the stationary state of the system
approaches a specific stationary state in which the probabilities of
finding electron in the levels become equal and the maximum
probability of tunneling from the conduction band of the left dot to
the conduction band of the right dot becomes $1/3$. To sum up, in
this work we showed that tunneling can be either increased or
decreased depending on the intensity of the laser, but due to the
presence of the environment there is always a maximum for the
populations of the two conduction bands of both dots and a minimum
for the population of the valence band of the left dot. The stronger
the applied laser is the closer the population of the second dot
will be to this maximum. We also saw whether the external driving
mode is applied or not there is always an equilibrium which the
system and the environment reach as $t\rightarrow\infty$.
\section{Appendix: A}
In the following we obtain the exact solutions of Eqs.(\ref{27}), (\ref{28}) and (\ref{29}).
 Applying Laplace transform to these equations we have
\begin{equation}\label{49}
s\rho_{00}(s)-\rho_{00}(t=0)=l\rho_{11}(s)-m\rho_{00}(s),
\end{equation}
\begin{equation}\label{50}
s\rho_{11}(s)-\rho_{11}(t=0)=-(l+n)\rho_{00}(s)+m\rho_{11}(s)+n\rho_{22}(s),
\end{equation}
\begin{equation}\label{51}
s\rho_{22}(s)-\rho_{22}(t=0)=n\rho_{11}(s)-n\rho_{22}(s),
\end{equation}
\begin{equation}\label{52}
\rho_{00}(s)+\rho_{11}(s)+\rho_{22}(s)=\frac{1}{s},
\end{equation}
where the last equality comes from the completeness relation, $tr\rho=1$.
After doing some calculations we find that
\begin{equation}\label{53}
\rho_{00}(s)=\frac{A}{s}+\frac{B}{s-\lambda_{0}}+\frac{C}{s-\lambda_{1}},
\end{equation}
where
\begin{equation}\label{54}
\lambda_{0}=\frac{-(l+m+2n)-\sqrt{(l+m+2n)^2-4(ln+2mn)}}{2},
\end{equation}
\begin{equation}\label{55}
\lambda_{1}=\frac{-(l+m+2n)+\sqrt{(l+m+2n)^2-4(ln+2mn)}}{2}.
\end{equation}
Now applying $\mathcal{L}^{-1}$ to $\rho_{00}(s)$ gives
$\rho_{00}(t)$ as
\begin{equation}\label{56}
\rho_{00}(t)=A+B\exp(\lambda_{0}t)+C\exp(\lambda_{1}t),
\end{equation}
where A, B and C are constants. A is the stationary value of $\rho_{00}(t)$ when time goes to infinity.
Substituting $\rho_{00}(t)$ into Eqs.(\ref{27}) and $tr\rho=1$, one can easily obtain $\rho_{11}(t)$ and $\rho_{22}(t)$ as
\begin{equation}\label{57}
\rho_{11}(t)=\frac{mA}{l}+\frac{(\lambda_{0}+m)B}{l}\exp(\lambda_{0}t)+\frac{(\lambda_{1}+m)C}{l}\exp(\lambda_{1}t),
\end{equation}
\begin{eqnarray}\label{58}
\rho_{22}(t)&=&1-(A+\frac{mA}{l})-\big(B+\frac{(\lambda_{0}+m)B}{l}\big)\exp(\lambda_{0}t)\nonumber \\
&-&\big(C+\frac{(\lambda_{1}+m)C}{l}\big)\exp(\lambda_{1}t).
\end{eqnarray}
The stationary values of $\rho_{00}(t)$, $\rho_{11}(t)$ and
$\rho_{22}(t)$ can be obtained by setting $\dot{\rho}_{00}(t)$,
$\dot{\rho}_{11}(t)$ and $\dot{\rho}_{22}(t)$ equal to zero.
Therefore from Eq.(\ref{29}) we have
\begin{equation}\label{59}
\rho_{11}(t)=\rho_{22}(t).
\end{equation}
Substituting Eqs.(\ref{57}) and (\ref{58}) into Eq.(\ref{59}) one gets
\begin{equation}\label{60}
A=\frac{l}{2m+l},
\end{equation}
which is the stationary value of $\rho_{00}(t)$. From the initial condition, $\rho_{00}(t=0)=1$ and $\rho_{11}(t=0)=0$, we find that
\begin{equation}\label{61}
B=1-\frac{l\big(m+(2m+l)(\lambda_{0}-\lambda_{1})\big)-2m(\lambda_{0}+m)}{(2m+l)(\lambda_{0}-\lambda_{1})}
\end{equation}
and
\begin{equation}\label{62}
C=\frac{m\big(l+2(\lambda_{0}+m)\big)}{(2m+l)(\lambda_{0}-\lambda_{1})}.
\end{equation}
Substituting A into Eqs.(\ref{57}) and (\ref{58}) the stationary values of both $\rho_{11}(t)$ and $\rho_{22}(t)$ are obtained to be $m/2m+l$.
Now consider the effect of the presence of an external resonant optical pulse in the stationary values of $\rho_{00}(t)$, $\rho_{11}(t)$ and $\rho_{22}(t)$. In this case there are $18$ coupled complex equations but all the matrix elements of the density matrix vanish except real parts of $\rho_{00}(t)$, $\rho_{11}(t)$ and $\rho_{22}(t)$ and imaginary parts of $\rho_{12}(t)$ and $\rho_{21}(t)$. Therefore we have five equations plus the completeness equation,
\begin{equation}\label{63}
\dot{\rho}_{00}(t)=l\rho_{11}(t)-m\rho_{00}(t)+p\rho_{21}(t)-p\rho_{12}(t),
\end{equation}
\begin{equation}\label{64}
\dot{\rho}_{11}(t)=-(l+n)\rho_{00}(t)+m\rho_{11}(t)+n\rho_{22}(t)-p\rho_{21}(t)+p\rho_{12}(t),
\end{equation}
\begin{equation}\label{65}
\dot{\rho}_{22}(t)=n\rho_{11}(t)-n\rho_{22}(t),
\end{equation}
\begin{equation}\label{66}
\dot{\rho}_{12}(t)=-\frac{1}{2}(l+m+n)\rho_{12}(t)-p\rho_{11}(t)+p\rho_{00}(t),
\end{equation}
\begin{equation}\label{67}
\dot{\rho}_{21}(t)=-\frac{1}{2}(m+n)\rho_{12}(t)+p\rho_{11}(t)-p\rho_{00}(t),
\end{equation}
\begin{equation}\label{68}
\rho_{00}(t)+\rho_{11}(t)+\rho_{22}(t)=1.
\end{equation}
Setting $\dot{\rho}_{00}(t)$, $\dot{\rho}_{11}(t)$, $\dot{\rho}_{22}(t)$, $\dot{\rho}_{12}(t)$ and $\dot{\rho}_{21}(t)$ equal to zero, after some calculations, the stationary values of $\rho_{00}(t)$, $\rho_{11}(t)$ and $\rho_{22}(t)$ can be easily obtained,
\begin{equation}\label{69}
\nonumber
\rho_{00}(t\rightarrow\infty)=\frac{l+2p^2(\frac{1}{m+n}+\frac{1}{l+m+n})}{2l+m+6p^2(\frac{1}{m+n}+\frac{1}{l+m+n})},
\end{equation}

\begin{equation}\label{70}
\nonumber
\rho_{11}(t\rightarrow\infty)=\rho_{22}(t\rightarrow\infty)=
1-\frac{2\big(l+2p^2(\frac{1}{m+n}+\frac{1}{l+m+n})\big)}{2l+m+6p^2(\frac{1}{m+n}+\frac{1}{l+m+n})}.
\end{equation}
As can be seen from  above equations when $p$ gets bigger, these
stationary values approach $1/3$ and in the limit of
$p\rightarrow\infty$  the three stationary values become $1/3$.

\newpage
\begin{figure}
\centering
\includegraphics[width=3.4cm]{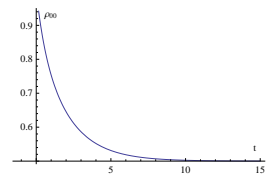}
\caption{$\rho_{00}(t)$ when $l=0.8$, $m=0.4$ and $\rho_{00}(t=0)=1$. As time goes to infinity $\rho_{00}(t)$ approaches a constant, $l/(2m+l)$.}
\label{fig:Fig2}
\end{figure}
\begin{figure}
\centering
\includegraphics[width=3.4cm]{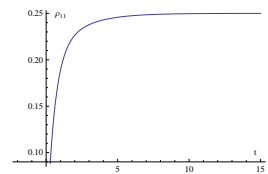}
\caption{$\rho_{11}(t)$ when $l=0.8$, $m=0.4$ and $\rho_{11}(t=0)=0$. As time goes to infinity $\rho_{11}(t)$ approaches a constant, $m/(2m+l)$.}
\label{fig:Fig3}
\end{figure}
\begin{figure}
\centering
\includegraphics[width=3.4cm]{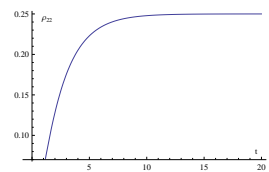}
\caption{$\rho_{22}(t)$ when $l=0.8$, $m=0.4$ and $\rho_{22}(t=0)=0$. As time goes to infinity $\rho_{22}(t)$ approaches a constant, $m/(2m+l)$.}
\label{fig:Fig4}
\end{figure}
\begin{figure}
\centering
\includegraphics[width=3.4cm]{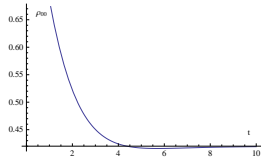}
\caption{$\rho_{00}(t)$ when $l/p=2$, $l=0.8$, $m=0.4$ and $\rho_{00}(t=0)=1$. $\rho_{00}(t)$ has a slight oscillation. As time passes
$\rho_{00}(t)$ approaches a constant.} \label{fig:Fig5}
\end{figure}
\begin{figure}
\centering
\includegraphics[width=3.4cm]{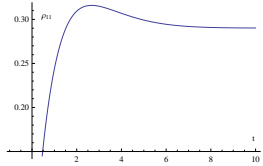}
\caption{$\rho_{11}(t)$ when $l/p=2$, $l=0.8$, $m=0.4$ and $\rho_{11}(t=0)=0$. $\rho_{11}(t)$ has a slight oscillation. As time passes $\rho_{11}(t)$ approaches a constant.}
\label{fig:Fig6}
\end{figure}
\begin{figure}
\centering
\includegraphics[width=3.4cm]{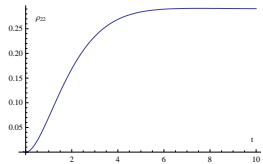}
\caption{$\rho_{22}(t)$ when $l/p=2$, $l=0.8$, $m=0.4$ and $\rho_{22}(t=0)=0$. As time passes $\rho_{22}(t)$ increases and then approaches a constant.}
\label{fig:Fig7}
\end{figure}
\begin{figure}
\centering
\includegraphics[width=3.4cm]{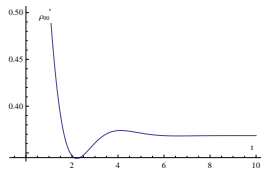}
\caption{$\rho_{00}(t)$ when $l/p=1$, $l=0.8$, $m=0.4$ and $\rho_{00}(t=0)=1$. $\rho_{00}(t)$ begins to oscillate. As time passes the number of oscillations of $\rho_{00}(t)$ decreases and then $\rho_{00}(t)$ approaches a constant.}
\label{fig:Fig8}
\end{figure}
\begin{figure}
\centering
\includegraphics[width=3.4cm]{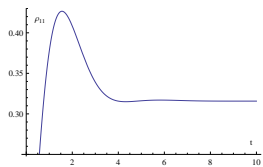}
\caption{$\rho_{11}(t)$ when $l/p=1$, $l=0.8$, $m=0.4$ and $\rho_{11}(t=0)=0$. $\rho_{11}(t)$ begins to oscillate. As time passes the number of oscillations of $\rho_{11}(t)$ decreases and then $\rho_{11}(t)$ approaches a constant.}
\label{fig:Fig9}
\end{figure}
\begin{figure}
\centering
\includegraphics[width=3.4cm]{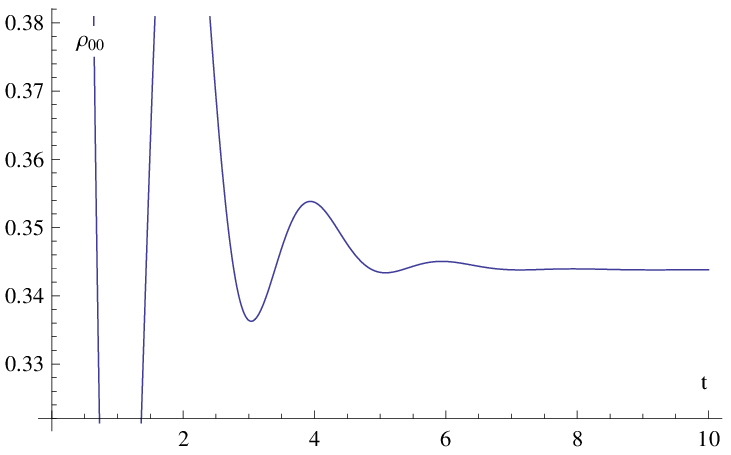}
\caption{$\rho_{00}(t)$ when $l/p=1/2$, $l=0.8$, $m=0.4$ and $\rho_{00}(t=0)=1$. As time passes the number of oscillations of $\rho_{00}(t)$ decreases and then $\rho_{00}(t)$ approaches a constant.}
\label{fig:Fig10}
\end{figure}
\begin{figure}
\centering
\includegraphics[width=3.4cm]{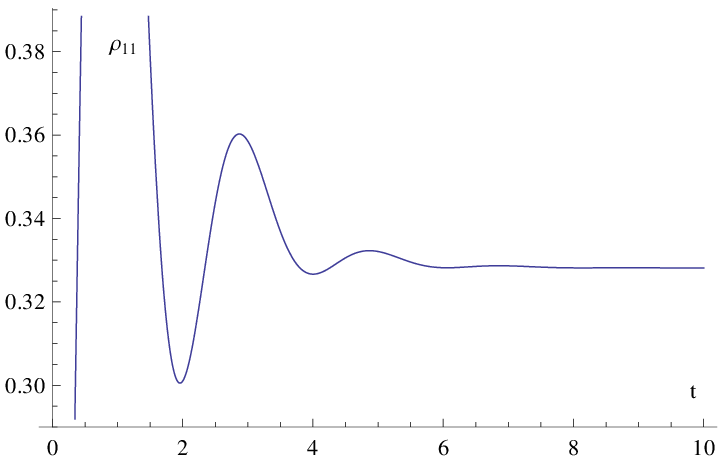}
\caption{$\rho_{11}(t)$ when $l/p=1/2$, $l=0.8$, $m=0.4$ and $\rho_{11}(t=0)=0$. As time passes the number of oscillations of $\rho_{11}(t)$ decreases and then $\rho_{11}(t)$ approaches a constant.}
\label{fig:Fig11}
\end{figure}
\begin{figure}
\centering
\includegraphics[width=3.4cm]{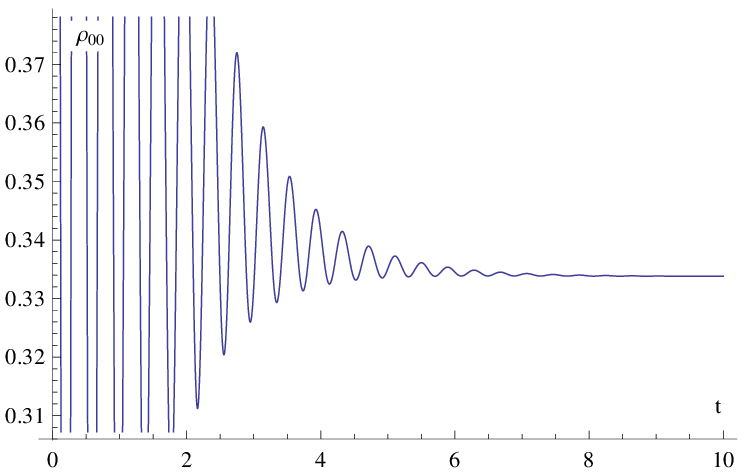}
\caption{$\rho_{00}(t)$ when $l/p=1/10$, $l=0.8$, $m=0.4$ and $\rho_{00}(t=0)=1$. As time passes the number of oscillations of $\rho_{00}(t)$ decreases and then $\rho_{00}(t)$ approaches a constant.}
\label{fig:Fig12}
\end{figure}
\begin{figure}
\centering
\includegraphics[width=3.4cm]{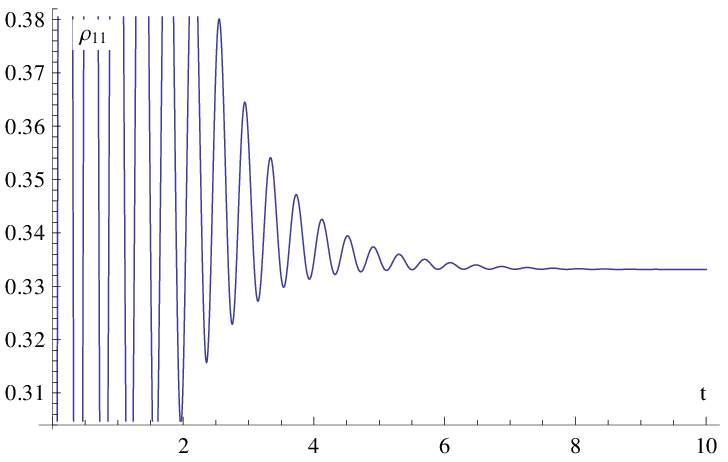}
\caption{$\rho_{11}(t)$ when $l/p=1/10$, $l=0.8$, $m=0.4$ and $\rho_{11}(t=0)=0$. As time passes the number of oscillations of $\rho_{11}(t)$ decreases and then $\rho_{11}(t)$ approaches a constant.}
\label{fig:Fig13}
\end{figure}
\end{document}